\documentclass[12pt]{iopart}

\usepackage{graphicx}
\usepackage{subfigure}
\usepackage{amsfonts}
\usepackage{amssymb}
\usepackage{bm}
\usepackage{color}
\usepackage[hypertex]{hyperref}

\begin{document}

  \title[H Meng {\it et al}]{Magnetism induced by nonlocal spin-entangled electrons in a superconducting spin-valve}
  \author{Hao Meng$^{1,2,3}$, Jiansheng Wu$^{1,4}$, Xiuqiang Wu$^{5}$, Mengyuan Ren$^6$, Yajie Ren$^2$ and Jinbin Yao$^2$}

  \address{$^1$Department of Physics and Institute for Quantum Science and Engineering, Southern University of Science and Technology, Shenzhen 518055, PR China}
  \address{$^2$School of Physics and Telecommunication Engineering, Shaanxi University of Technology, Hanzhong 723001, PR China}
  \address{$^3$Shanghai Key Laboratory of High Temperature Superconductors, Shanghai University, Shanghai 200444, PR China}
  \address{$^4$Shenzhen Key Laboratory of Quantum Science and Engineering, Shenzhen 518055, PR China}
  \address{$^5$School of Mathematics and Physics, Yancheng Institute of Technology, Yancheng 224051, PR China}
  \address{$^6$College of materials science and engineering, Fuzhou University, Fuzhou 350116, PR China}
  \date{\today}
  \ead{wu.js@sustc.edu.cn}

   \begin{abstract}
   In the traditional view, the magnetic moment appearing in the superconducting region is induced by equal-spin triplet superconducting correlations in superconductor ($S$) ferromagnet ($F$) heterostructure with noncollinear magnetization. In this paper, we represent that in $NSF_1F_2$ ($N$--normal-metal) spin-valve structure the induced magnetic moment emerging in both the $S$ and $N$ regions can also be generated by Cooper pair splitting: one electron coherently tunnels from the $S$ layer into the $F_1$ layer, and the other one stays in the $S$ layer or tunnels into the $N$ layer. Two electrons are spatially separated from each other but their total spin ground state is entangled in this process. In contrast, the magnetic moment induced by the equal-spin triplet correlations hardly penetrates from the $S$ layer into the $N$ layer. In particular, by tuning the size of the exchange field and the thickness of the $F_1$ layer, one may control the direction of the induced magnetic moment in the $N$ layer. This interesting phenomenon can be attributed to the phase-shift obtained by the spin-entangled electrons. Our theoretical proposal will offer an effective way to control the entanglement of the nonlocal electrons, and also may provide possible explanations for previous and recent experimental observations [Stamopoulos \emph{et al} 2005 {\it Phys. Rev. B} {\bf 72} 212514; Ovsyannikov \emph{et al} 2016 {\it J. Exp. Theor. Phys.} {\bf 122} 738; Flokstra \emph{et al} 2016 {\it Nat. Phys.} {\bf 12} 57].

   \end{abstract}


   \section{Introduction}

  The interplay between superconductivity and ferromagnetism in hybrid structures has currently attracted considerable attention because of unusual physical phenomena~\cite{ZoharNussinov,IverBSperstad,MadalinaColci,KueiSun} and potential practical applications~\cite{AAGolubov,Buz,BerRMP,Esc,MatthiasEschrig}. Various device applications of superconducting hybrid structures have been widely studied in recent years~\cite{SolovievII,BhattiS,YamashitaT}. Much effort has been devoted to obtaining a better understanding of these phenomena appeared in heterostructures involving superconductor ($S$) and ferromagnet ($F$). It is well known that ferromagnetism and conventional superconductivity are two antagonistic orders, as ferromagnetism favors a parallel spin alignment, while Cooper pairs consist of electrons with antiparallel aligned spins. The interaction of these two orders leads to the proximity effect and the inverse proximity effect.

  The proximity effect causes the superconducting correlations to penetrate into the ferromagnetic region. When the superconductor is adjacent to a homogeneous ferromagnet, the Cooper pairs entering the ferromagnetic region acquire a finite center-of-mass momentum $Q\simeq2h_0/\hbar{v_F}$ due to the exchange splitting of the ferromagnet. Here $\emph{h}_0$ and $v_F$ are the exchange field strength and the Fermi velocity, respectively. The wave function of the Cooper pair oscillates and decays in the ferromagnetic region as a function of $QR$, where $R$ is the distance from the $S/F$ interface~\cite{Buz,BerRMP,Esc,MatthiasEschrig}. This oscillatory nature leads to periodic $\pi$-phase shifts across the $SFS$ junction~\cite{FSBAFVKBE}. Another unusual effect highlighted in the $SF$ heterostructures is the fact that inhomogeneous textures of the magnetization in the ferromagnet may lead to the creation of spin triplet superconducting correlations with equal spin projections on the quantization axis~\cite{Esc,MatthiasEschrig,MEschrigTL}. These triplet correlations are not destroyed by the exchange field and can propagate a long distance in the strongly spin-polarized ferromagnet and half-metal. The experimental evidence of such triplet correlations is revealed by the recent observations of the long-range Josephson current~\cite{TruptiSKhaire,CarolinKlose,JWARobinson,RSKeizer,MSAnwar} and the superconducting transition temperature in the $S/F$ spin-valve heterostructure~\cite{VIZdravkov,PVLeksin,XLWang,AAJara}.

  In contrast, the inverse proximity effect induces ferromagnetic correlation in the superconducting region near the $S/F$ interface. It was proposed theoretically that in a $SF$ bilayer the magnetic moment could be induced in the superconducting region and its direction is opposite to the magnetization in the bulk of the ferromagnet~\cite{FSBAFV,JXVSMK}. This induced magnetic moment also displays an oscillatory sign-changing behavior changing with the product $h_0d$ of the ferromagnetic thickness $d$~\cite{MYKAFVKBE,JLTY}. If the magnetization distribution is inhomogeneous in the ferromagnetic region, then the equal-spin triplet correlations give rise to an induced magnetic moment in the superconducting layer~\cite{Krivoruchko,TLTCJDME,NGPAIB}. Previously, Stamopoulos \emph{et al.}~\cite{DStaNMou} offered an experimental result in the S/manganite multilayers/F structures. They observed that the inhomogeneous magnetization structure of the manganite multilayers, which was modulated by the external magnetic field, could efficiently switch the direction of the magnetic moment appearing in the superconducting region. Recently, Ovsyannikov \emph{et al.}~\cite{GAOvsyannikov} reported experimentally that the magnetic moment emerged in the superconducting part of the heterostructure that consists of a cuprate superconductor and a ferromagnetic spin valve when the magnetization vectors of the ferromagnetic films have noncollinear orientation.

  On the other hand, it is known that the conventional superconductor in principle is considered as a natural source of spin entanglement~\cite{PRecher,GBLesovik}. When two spatially separated normal-metal electrodes form two separate junctions with a superconductor, with the junction separation of the order of the superconducting coherence length of the material, the Cooper pair in the superconductor can break up into two nonlocal entangled electrons that enter into different electrodes via so-called the crossed Andreev reflection or Cooper pair splitting (CPS)~\cite{JSchindele}. This effect  can achieve the nonlocal electronic correlations ~\cite{DBeckmann,SRusso,PCaddenZimansky,AKleine,JWei,DanielLoss}: two entangled electrons are localized in the different normal-metal electrodes, and while being spatially separated from each other. What happens if one of the normal-metal electrodes is replaced by an inhomogeneous ferromagnet? Can the electronic spin in the normal-metal electrode be dominated by the characteristics of the ferromagnet? Based on these questions, Flokstra \emph{et al.}~\cite{MGFNSJKGB} represented a surprising experimental observation in $NSF_1F_2$ multilayered structure. The induced magnetic moment appears inside the normal-metal, but not in the adjacent superconducting layer. In particular, the magnetic moment exhibits a spin-valve effect: a significant change in magnitude depends on the mutual orientation of magnetization in two ferromagnetic layers. Nevertheless, a key issue still needs to be resolved and further studied: which component, the spin singlet correlation or the equal-spin triplet correlation, makes a significant contribution to the magnetic moment induced in the normal-metal?

  In this paper, we demonstrate that the CPS effect could induce a magnetic moment inside the superconductor and the normal-metal in the $NSF_1F_2$ structure with noncollinear magnetization. The equal-spin triplet pairs within the superconducting region also produce a  magnetic moment in the different direction, but this magnetic moment hardly penetrates from superconductor to normal-metal. In this spin-valve structure, the induced magnetic moment in both the superconductor and the normal-metal is different from the homogenous $SF$ structure, in which the magnetic moment occurs in the superconducting region only when the exchange field of the ferromagnet is weak enough. However, in our structure the induced magnetic moment still exists in the superconductor and the normal-metal even if the $F_2$ layer converts into a strongly spin-polarized ferromagnet. In particular, the direction of the induced magnetic moment could be reversed by changing the exchange field and the thickness of the $F_1$ layer. This feature can be attributed to the phase-shift obtained by the entangled electrons. However, the change of the $F_2$ layer triggers an amplitude oscillation of the induced magnetic moment inside the $N$ region but does not reverse its direction. When the exchange field or the thickness of the $F_2$ layer becomes large, the oscillation will decrease accordingly. Moreover, the induced magnetic moment depends on the temperature and the misorientation angle between the two ferromagnetic layers. This effect can be used for engineering the cryoelectronic devices manipulating the magnetic moment in the normal-metal.

   \section{Model and formula}

\begin{figure}
    \centerline{\includegraphics[width=3.3in]{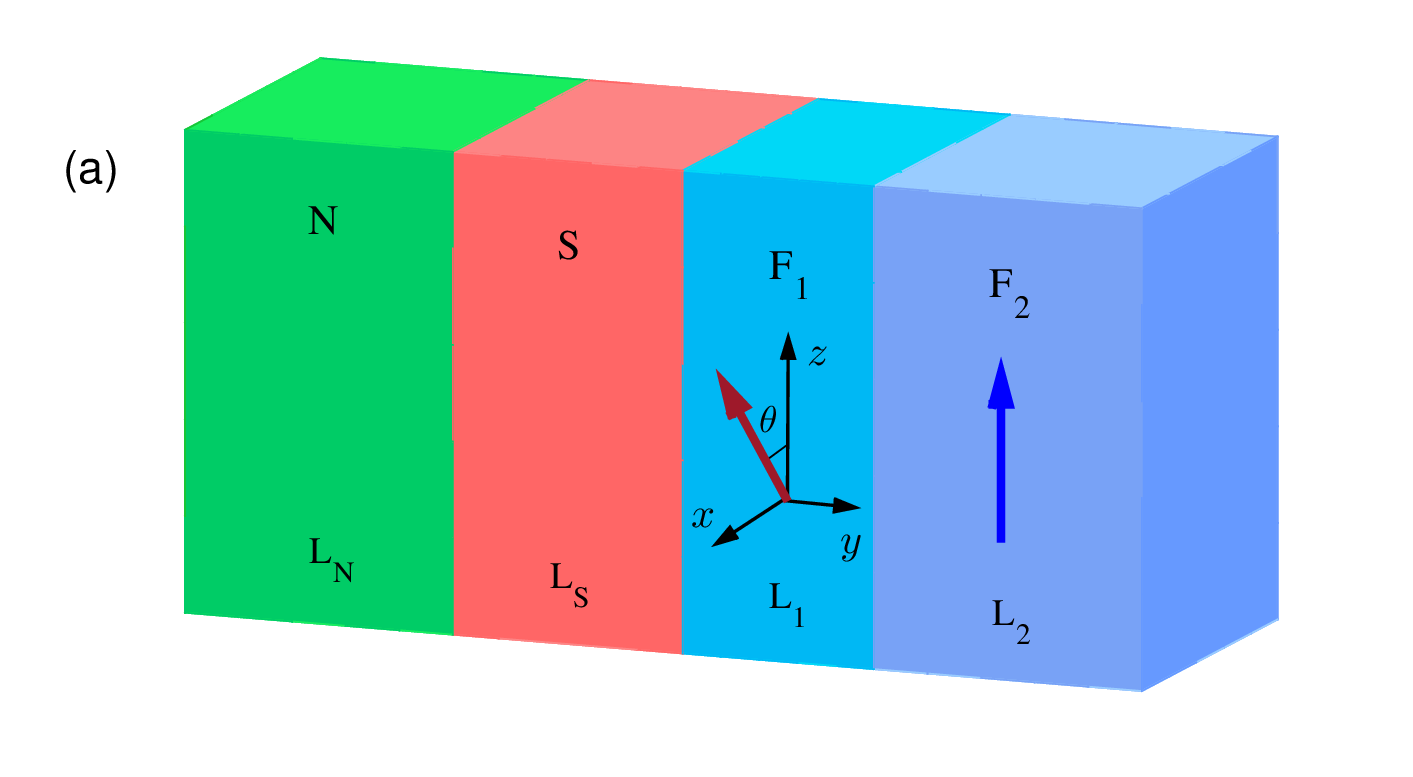}}
    \centerline{\includegraphics[width=3.35in]{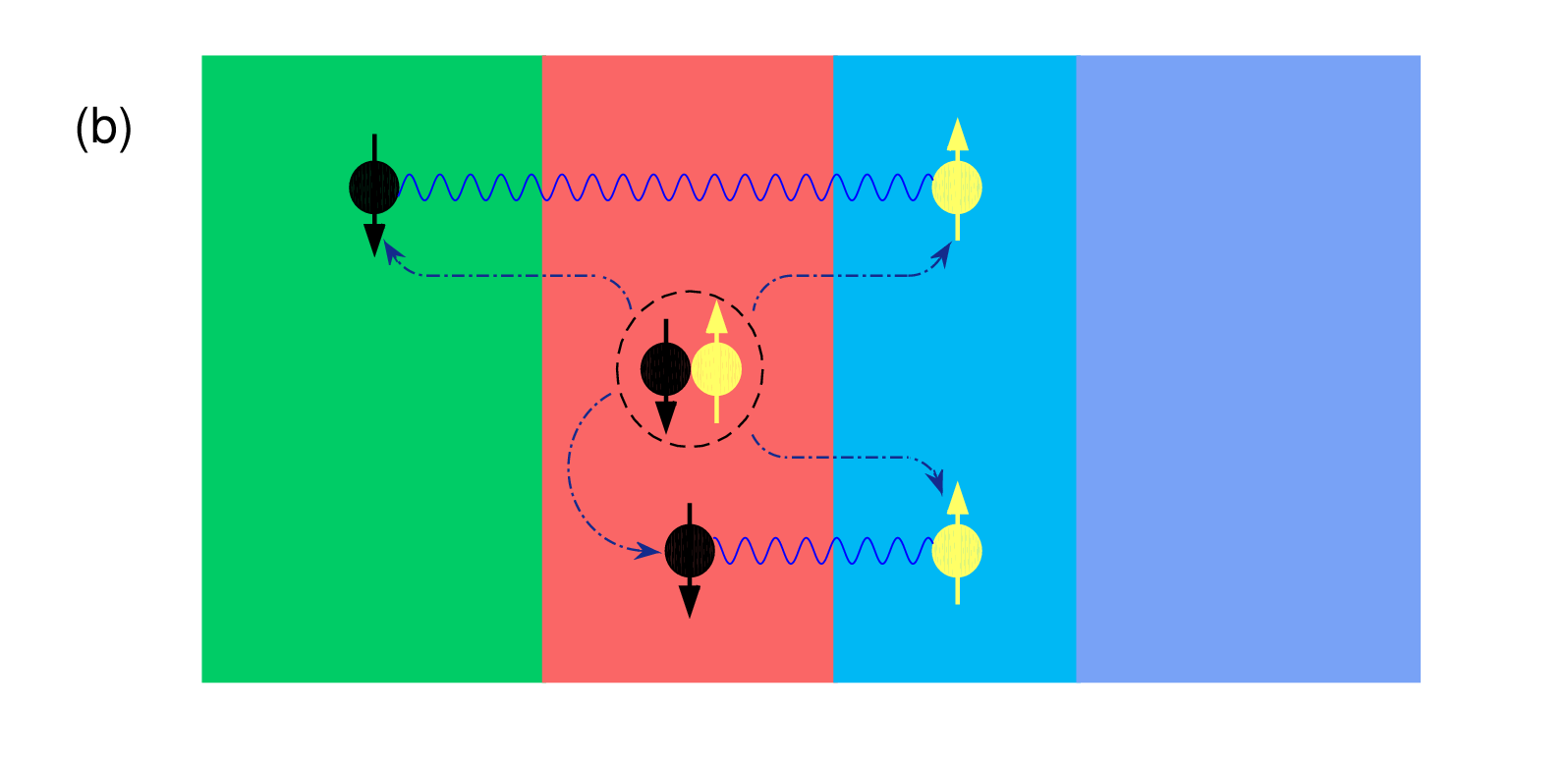}}
    \caption{(a) Schematic of $NSF_1F_2$ structure. Thick arrows indicate the directions of the magnetization in two ferromagnetic layers. All layer widths are labeled. (b) Schematic of Cooper pair splitting: two spin-entangled electrons forming a Cooper pair tunnel from $S$ to $N$ and $F_1$, or one electron stays in $S$ and the other one tunnels into $F_1$.}
    \label{Fig.1}
    \end{figure}

    The $NSF_1F_2$ junction we consider is shown schematically in figure~\ref{Fig.1}(a). We denote the layer thicknesses by $L_N$, $L_S$, $L_1$, and $L_2$, respectively. The origin of the $y$ axis, that is perpendicular to the layer interfaces, locates at the outer $N$ surface. The magnetization of the $F_1$ layer is tilted by misorientation angle $\theta$ from the $z$-axis in the $x$-$z$ plane, while the magnetization of the $F_2$ layer is aligned along the $z$-axis. We also assume that the whole system satisfies translational invariance in the $x$-$z$ plane.

    The BCS mean-field effective Hamiltonian~\cite{Buz,PGdeGennes} is
    \begin{eqnarray}
     H_{eff}&=\sum_{\alpha,\beta}\int{d\vec{r}}\{\psi^{\dag}_{\alpha}(\vec{r})H_e\psi_{\alpha}(\vec{r})-\psi^{\dag}_{\alpha}(\vec{r})(\vec{h}\cdot\vec{\sigma})_{\alpha\beta}\psi_{\beta}(\vec{r}) \nonumber\\
     &+\frac{1}{2}[(i\sigma_{y})_{\alpha\beta}\Delta(\vec{r})\psi^{\dag}_{\alpha}(\vec{r})\psi^{\dag}_{\beta}(\vec{r})+h.c.]\},
     \label{Heff}
    \end{eqnarray}
    where $H_e=-\hbar^{2}\nabla^{2}/2m-E_F$, $\psi^{\dag}_{\alpha}(\vec{r})$ and $\psi_{\alpha}(\vec{r})$ represent creation and annihilation operators with spin $\alpha$, and the vector $\vec{\sigma}=(\sigma_x, \sigma_y, \sigma_z)$ is composed of Pauli spin matrices. $m$ is the effective mass of the quasiparticles in the system, and $E_F$ is the Fermi energy. $\Delta(\vec{r})$ describes the usual superconducting pair potential. The exchange field exists only in the ferromagnetic region. It can be described by
    \begin{equation*}
    \vec{h}=
    \cases{h_1(\sin\theta\hat{x}+\cos\theta\hat{z}), & $L_{N}+L_{S}<y<L_{N}+L_{S}+L_{1}$, \\
    h_2\hat{z}, & $L_{N}+L_{S}+L_{1}<y<L_{N}+L_{S}+L_{1}+L_{2}$, \\}
   \end{equation*}
   where $\hat{x}$ ($\hat{z}$) is the unit vector along the $x$ ($z$) direction.

  In order to diagonalize the effective Hamiltonian, we make use of the Bogoliubov transformation $\psi_{\alpha}(\vec{r})$=$\sum_{n}[u_{n\alpha}(y)\hat{\gamma}_{n}$+$v^{\ast}_{n\alpha}(y)\hat{\gamma}^{\dag}_{n}]$ and take into account the anticommutation relations of the quasiparticle annihilation operator $\hat{\gamma}_{n}$ and creation operator $\hat{\gamma}^{\dag}_{n}$. The resulting Bogolubov-de-Gennes (BdG) equation can be expressed as~\cite{PGdeGennes}

  \begin{equation}
  \left(
  \begin{array}{ccc}
	\hat{H}(y) & i\hat{\sigma}_{y}\Delta(y) \\
    -i\hat{\sigma}_{y}\Delta^{\ast}(y) & -\hat{H}(y) \\
  \end{array}
  \right)
  \left(
  \begin{array}{ccc}
	\hat{u}_n(y) \\
    \hat{v}_n(y) \\
  \end{array}
  \right)
  =E_n
  \left(
  \begin{array}{ccc}
	\hat{u}_n(y) \\
    \hat{v}_n(y) \\
  \end{array}
  \right),
  \label{BdG}
  \end{equation}
  where $\hat{H}(y)=H_e\hat{\textbf{1}}-h_z(y)\hat{\sigma}_{z}-h_{x}(y)\hat{\sigma}_{x}$ and $\hat{\textbf{1}}$ is the unity matrix. Besides, $\hat{u}_n(y)=[u_{n\uparrow}(y), u_{n\downarrow}(y)]^T$ and $\hat{v}_n(y)=[v_{n\uparrow}(y), v_{n\downarrow}(y)]^{T}$ are quasiparticle and quasihole wave functions, respectively.

  To acquire the characteristics of the physical quantities in the system, we solve the BdG equation~(\ref{BdG}) by Bogoliubov's self-consistent field method~\cite{PGdeGennes,JBKetterson,KlausHalterman,HaoMeng}. The $NSF_1F_2$ junction is placed in a one-dimensional square potential well with infinitely high walls, then the corresponding quasiparticle amplitudes can be expanded in terms of a set of basis vectors of the stationary states~\cite{LDLandau}, $u_{n\alpha}(y)=\sum_{p}u^{\alpha}_{np}\zeta_{p}(y)$ and $v_{n\alpha}(y)=\sum_{p}v^{\alpha}_{np}\zeta_{p}(y)$ with $\zeta_{p}(y)=\sqrt{2/L}\sin(p{\pi}y/L)$. Here $p$ is a positive integer and $L=L_{N}+L_S+L_{1}+L_{2}$. The pair potential in the BdG equation~(\ref{BdG}) satisfies the self-consistency condition~\cite{PGdeGennes}

  \begin{equation}
    \Delta(y)=\frac{g(y)}{2}\sum_{n}{'}\sum_{pp'}(u_{np}^{\uparrow}v_{np'}^{\downarrow*}-u_{np}^{\downarrow}v^{\uparrow*}_{np'})\zeta_{p}(y)\zeta_{p'}(y)\tanh(\frac{E_n}{2k_BT}),
    \label{Det}
  \end{equation}
  where $k_B$ and $T$ are the Boltzmann constant and the temperature, respectively. The superconducting oupling parameter $g(y)$ will be taken to be constant in the superconducting region and zero elsewhere. The primed sum is over the eigenstates within range $|E_n|\leq\omega_{D}$, where $\omega_{D}$ is the Debye cutoff energy. The matrix elements in equation~(\ref{BdG}) are then written as
  \begin{equation}
    H_e(q,q')=\int_{0}^{L}\zeta_{q}(y)[-\frac{1}{2m}\frac{\partial^{2}}{\partial{y}^{2}}+\varepsilon_{\bot}-E_F]\zeta_{q'}(y)dy,
    \label{EqHe}
  \end{equation}
  \begin{equation}
    h_x(q,q')=h_1\sin\theta\int_{L_{N}+L_{S}}^{L_{N}+L_{S}+L_{1}}\zeta_{q}(y)\zeta_{q'}(y)dy,
    \label{Eqhx}
  \end{equation}
  \begin{equation}
    h_z(q,q')=h_1\cos\theta\int_{L_{N}+L_{S}}^{L_{N}+L_{S}+L_{1}}\zeta_{q}(y)\zeta_{q'}(y)dy+h_2\int_{L_{N}+L_{S}+L_{1}}^{L}\zeta_{q}(y)\zeta_{q'}(y)dy,
    \label{Eqhz}
  \end{equation}
  \begin{equation}
    \Delta(q,q')=\frac{1}{2}\sum_{n}{'}\sum_{pp'}(u_{np}^{\uparrow}v_{np'}^{\downarrow*}-u_{np}^{\downarrow}v^{\uparrow*}_{np'})\tanh(\frac{E_n}{2k_BT})\int_{0}^{L}g(y)\zeta_{p}(y)\zeta_{p'}(y)\zeta_{q}(y)\zeta_{q'}(y)dy,
    \label{EqDelta}
  \end{equation}
  where $\varepsilon_{\bot}$ in equation~(\ref{EqHe}) is the continuous energy in the transverse direction. The BdG equation~(\ref{BdG}) is solved by an iterative schedule. One first starts from the stepwise approximation for the pair potential and iterations are performed until the change in value obtained for $\Delta(y)$ does not exceed a small threshold value.

  The local magnetic moment of the $NSF_1F_2$ geometry has two components~\cite{KlausHalterman}
  \begin{eqnarray}
   m_{x}(y)&=-\mu_{B}\sum_{n}\sum_{qq'}[(u^{\uparrow*}_{nq}u^{\downarrow}_{nq'}+u^{\downarrow*}_{nq}u^{\uparrow}_{nq'})f_n \nonumber\\
   &+(v^{\uparrow}_{nq}v^{\downarrow*}_{nq'}+v^{\downarrow}_{nq}v^{\uparrow*}_{nq'})(1-f_n)]\zeta_{q}(y)\zeta_{q'}(y),
   \label{Eq9}
  \end{eqnarray}
  \begin{eqnarray}
   m_{z}(y)&=-\mu_{B}\sum_{n}\sum_{qq'}[(u^{\uparrow*}_{nq}u^{\uparrow}_{nq'}-u^{\downarrow*}_{nq}u^{\downarrow}_{nq'})f_n \nonumber\\
   &+(v^{\uparrow}_{nq}v^{\uparrow*}_{nq'}-v^{\downarrow}_{nq}v^{\downarrow*}_{nq'})(1-f_n)]\zeta_{q}(y)\zeta_{q'}(y),
   \label{Eq10}
  \end{eqnarray}
  where $\mu_{B}$ and $f_n$ are the effective Bohr magneton and the Fermi function, respectively. It is convenient to normalize these two components to $-\mu_{B}$. Because the local magnetic moment in the normal-metal is rather small, we define an effective magnetic moment $M^N_{x(z)}=\int_{0}^{L_{N}}m_{x(z)}(y)dy$ to describe the total effect of the magnetic moment induced in the normal-metal region.

  The amplitude functions of the spin triplet pair with zero spin projection and the equal-spin triplet pair are defined as follows~\cite{KlausHalterman}
  \begin{equation}
   f_{0}(y,t)=\frac{1}{2}\sum_{n}\sum_{qq'}(u^{\uparrow}_{nq}v^{\downarrow*}_{nq'}+u^{\downarrow}_{nq}v^{\uparrow*}_{nq'})\zeta_{q}(y)\zeta_{q'}(y)\eta_n(t),
  \label{F0}
  \end{equation}
  \begin{equation}
   f_{1}(y,t)=\frac{1}{2}\sum_{n}\sum_{qq'}(u_{nq}^{\uparrow}v^{\uparrow*}_{nq'}-u_{nq}^{\downarrow}v^{\downarrow*}_{nq})\zeta_{q}(y)\zeta_{q'}(y)\eta_n(t),
  \label{F1}
  \end{equation}
  where the sum of $E_n$ is in general performed over all positive energies, and $\eta_n(t)=\cos(E_nt)-i\sin(E_nt)\tanh(E_n/2k_BT)$. Additionally, the amplitude function of the spin singlet pair can be written as
  \begin{equation}
    f_{3}(y)=\frac{1}{2}\sum_{n}{'}\sum_{pp'}(u_{np}^{\uparrow}v_{np'}^{\downarrow*}-u_{np}^{\downarrow}v^{\uparrow*}_{np'})\zeta_{p}(y)\zeta_{p'}(y)\tanh(\frac{E_n}{2k_BT}),
    \label{f3}
  \end{equation}
  In this paper the singlet and triplet amplitude functions are all normalized to the value of the singlet pairing amplitude in a bulk superconducting material.

  The LDOS is obtained according to~\cite{KlausHalterman}
  \begin{eqnarray}
   N(y,\epsilon)&=-\sum_{n}{'}\sum_{qq'}[(u_{nq}^{\uparrow}u^{\uparrow*}_{nq'}+u_{nq}^{\downarrow}u^{\downarrow*}_{nq'})f'(\epsilon-E_n) \nonumber\\
   &+(v_{nq}^{\uparrow}v^{\uparrow*}_{nq'}+v_{nq}^{\downarrow}v^{\downarrow*}_{nq'})f'(\epsilon+E_n)]\zeta_{q}(y)\zeta_{q'}(y),
  \label{LDOS}
  \end{eqnarray}
  where $f'(\epsilon)=\partial{f}/\partial{\epsilon}$ is the derivative of the Fermi function. The amplitude of the LDOS is normalized by its value at $\epsilon=2\Delta$ beyond which the LDOS is almost constant.

  \section{Results and Discussions}

  In this section we present and discuss the results obtained by solving numerically the BdG equation (\ref{BdG}). All lengths and the exchange field strengths are measured in units of the inverse Fermi wave vector $k_F$ and the Fermi energy $E_F$, respectively. The superconducting coherence length and the Debye cutoff energy are taken as $k_F\xi_S$=100 and $\omega_{D}/E_F$=0.1, respectively. The $N$ layer thickness is assumed to be $k_FL_{N}$=70 and all interfaces in the system are fully transparent.

  \begin{figure}
  \centerline{\includegraphics[width=6.0in]{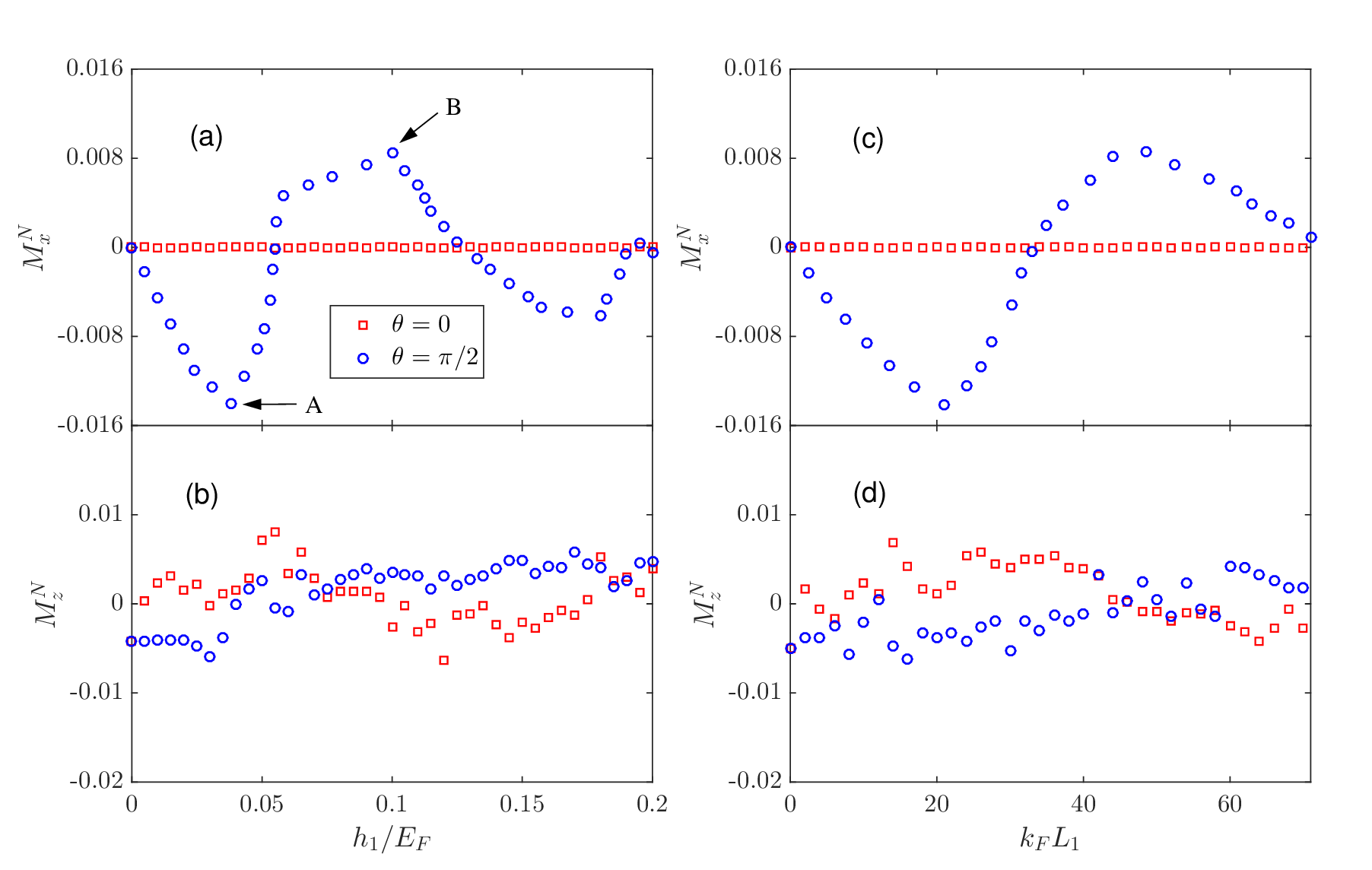}} 
  \caption{The effective magnetic moments $M^N_x$ (a) and $M^N_z$ (b) inside the $N$ region as a function of exchange field $h_1/E_F$ for thickness $k_FL_1=20$. The corresponding $M^N_x$ (c) and $M^N_z$ (d) as a function of thickness $k_FL_1$ for exchange field $h_1/E_F=0.035$. All results shown in this figure are for $k_FL_S=50$, $k_FL_2=50$, $h_2/E_F=0.2$, and $k_BT=0$.}
  \label{Fig.2}
  \end{figure}

  \begin{figure}
  \centerline{\includegraphics[width=6in]{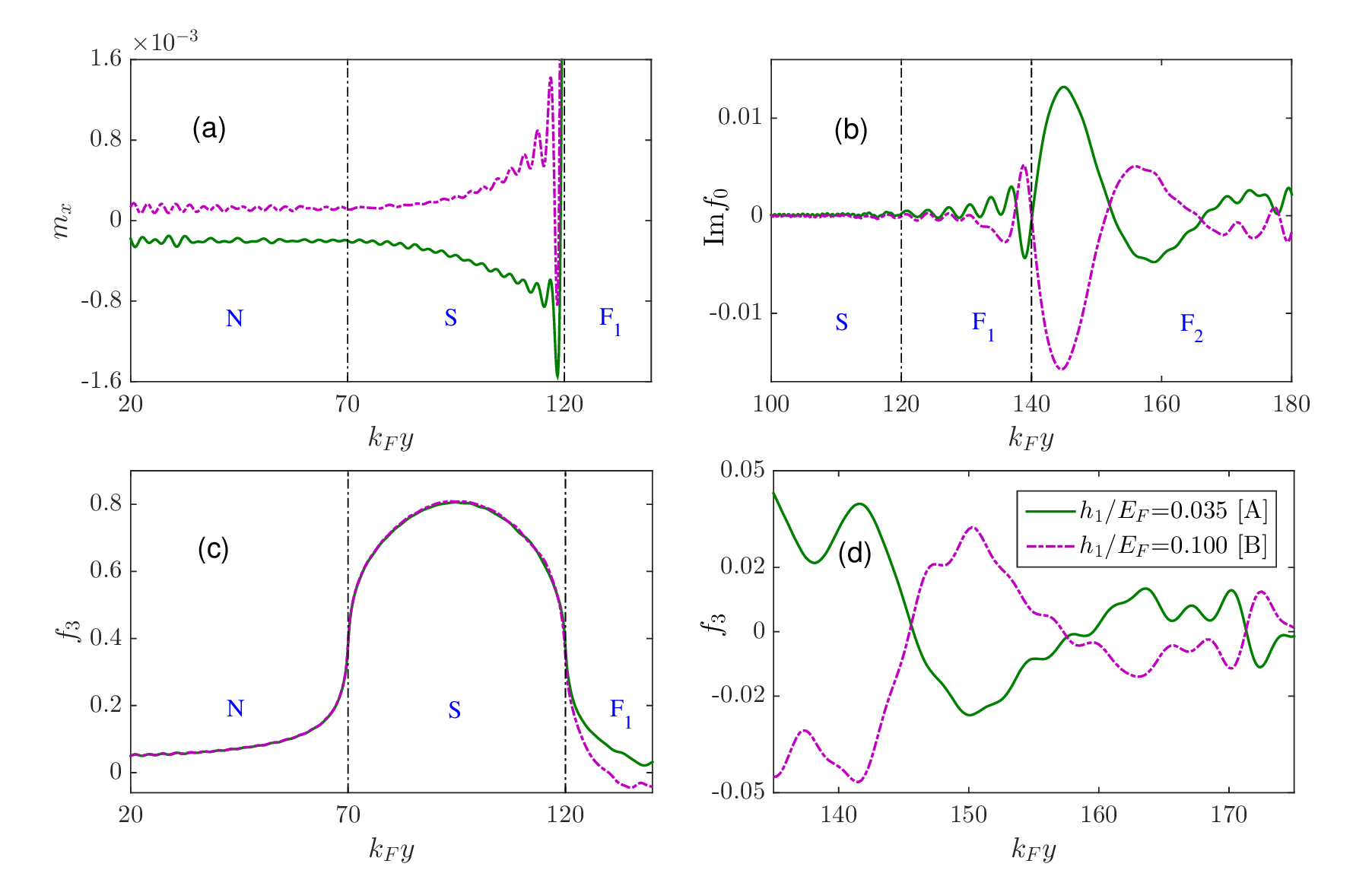}} 
  \caption{Spatial distribution of the induced magnetic moment $m_x$ (a), the imaginary parts of the spin triplet state $f_0$ (b), the spin singlet state $f_3$ (c) and the zoom of $f_3$ in the ferromagnetic region for $\theta=\pi/2$. The results plotted are for $k_FL_S=50$, $k_FL_1=20$, $k_FL_2=50$, $h_2/E_F=0.2$, and $k_BT=0$. Here $\omega_Dt=2$ is used in (b). The vertical dash-dotted lines represent the locations of the interface between the layers.}
  \label{Fig.3}
  \end{figure}

  In figure~\ref{Fig.2} we show the detailed dependence of the effective magnetic moments $M^N_x$ and $M^N_z$ inside the $N$ region on the exchange field $h_1$ and the thickness $L_1$. It is clear see that $M^N_x$ does not exist in the parallel configuration ($\theta$=0), but it displays an oscillating sign-changing behavior upon increasing $h_1$ or $L_1$ in the perpendicular configuration ($\theta$=0.5$\pi$). This indicates that the direction of $M^N_x$ can be reversed by varying the strength of $h_1$ or $L_1$. By contrast, $M^N_z$ changes irregularly and its magnitude is usually very small in the above two cases. Since the exchange field $h_1$ is hardly variable in the experiment, one may hope to observe the oscillations of $M^N_x$ performing measurements on samples with the different thickness $L_1$. It needs to emphasize that in figure~\ref{Fig.2} (and figure~\ref{Fig.8} depicted below) the intervals between data points have been taken to be large enough for the sake of beauty. In fact, there are other small oscillation peaks in the interval, which may be ascribed to other mechanisms. For simplicity, we ignore these additional oscillation peaks.

  In general, since the entangled electrons penetrate into the ferromagnetic region, their spins will be arranged along magnetization direction of the ferromagnet. When both ferromagnetic layers are all parallel to $z$-axis, $m_z$ will be induced in both the $S$ and $N$ regions for the weak exchange field. This behavior will be discussed in detail below. If the total exchange field of two ferromagnetic layers is chosen to be strong enough, which causes a depairing effect, it is difficult for the entangled electrons to tunnel into these ferromagnetic regions, then the CPS effect can hardly occur. Correspondingly, $m_z$ shows a quite small amplitude. When the magnetization directions of both ferromagnets are perpendicular to each other ($\theta$=$\pi/2$), the physical picture will become more complicated. Two effects will cause two magnetization components along different directions. Firstly, the equal-spin triplet pair generates a component $m_z$ in the $S$ region but not in the $N$ region. Secondly, the CPS effect produces the other component $m_x$ in both the $S$ and $N$ regions. The direction of $m_x$ will be reversed with varying characters of $F_1$ layer because the entangled electrons obtain an additional phase in this situation. In the following, we will demonstrate the above descriptions in detail using the numerical calculation results.

   \begin{figure}
   \centerline{\includegraphics[width=6.0in]{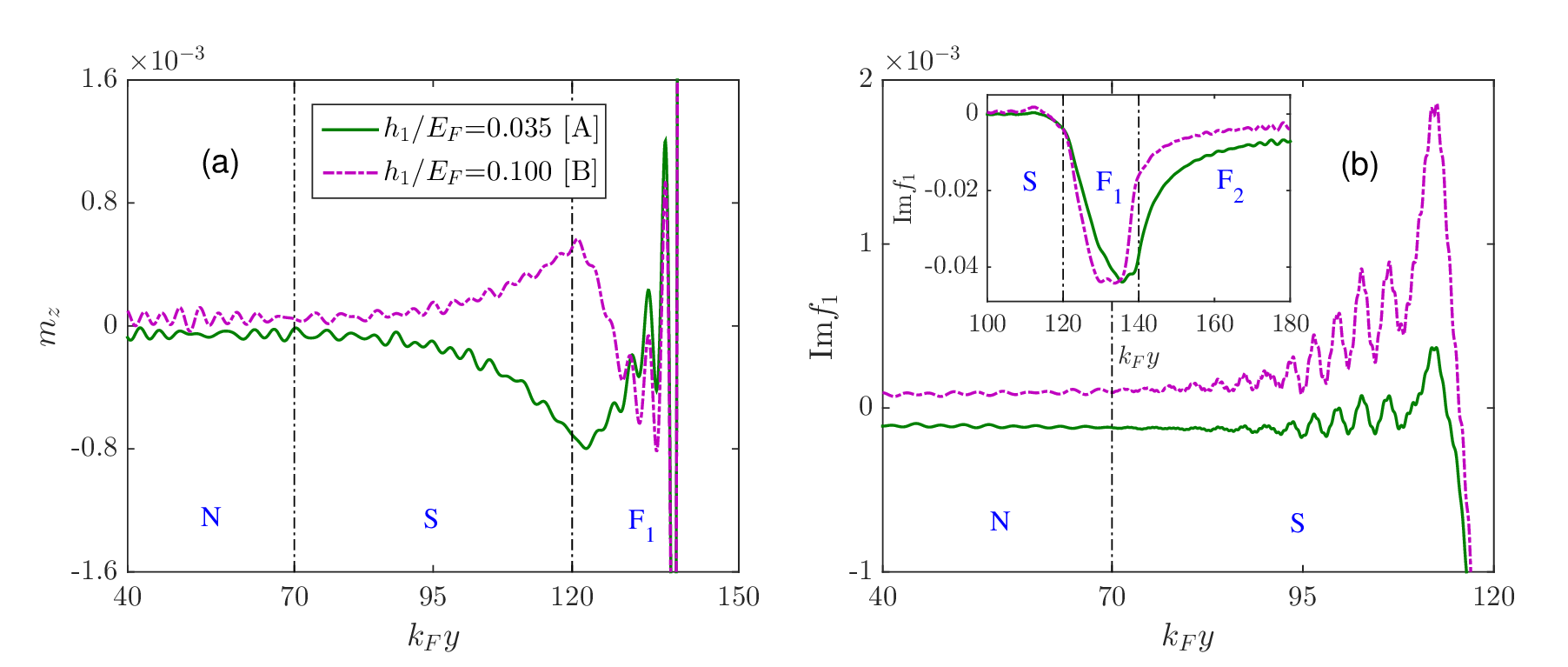}} 
   \caption{The spatial distribution of the induced magnetic moment $m_z$ (a) and the imaginary parts of the spin triplet state $f_1$ in the $N$ and $S$ regions (b) for two different exchange fields $h_1/E_F$. Inset in (b) shows $f_1$ in the $S$ and $F$ regions. Here $\omega_Dt=2$ is used in (b), and other parameters are the same as in figure~\ref{Fig.3}. The vertical dash-dotted lines indicate the interfaces between the layers.}
   \label{Fig.4}
   \end{figure}

  In figure~\ref{Fig.3} (a) we show the spatial profile of the induced magnetic moment $m_x$ for two special exchange fields $h_1/E_F$=0.035 and 0.1, which corresponds to the points A and B in figure~\ref{Fig.2} (a), respectively. In the former case ($h_1/E_F$=0.035), the negative $m_x$ decays gradually as a function of distance from the $S/F_1$ interface. When $m_x$ penetrates from the $S$ region into the $N$ region, its sign and magnitude almost remain unchanged inside the whole $N$ region. This phenomenon can be ascribed to the CPS effect~\cite{JSchindele}. As illustrated in figure~\ref{Fig.1} (b), two entangled electrons forming a Cooper pair transport from the $S$ region into the $N$ region and the $F_1$ region, or one electron stays in the $S$ region and the other one tunnels to the $F_1$ region. The accumulation of these electrons leads to the magnetic moment in the $N$ and $S$ regions, whose direction is opposite to the exchange field in the $F_1$ layer. In the latter case ($h_1/E_F$=0.035), the direction of $m_x$ may turn from negative to positive, which is due to the $\pi$ phase-shift acquired by the entangled electrons. The detailed analysis is described below: the spin singlet state $\mid\uparrow\downarrow\rangle$$-$$\mid\downarrow\uparrow\rangle$ originating from the $S$ layer can be converted into $\mid\uparrow\downarrow\rangle_x$$e^{iQ\cdot{R}}$$-$$\mid\downarrow\uparrow\rangle_x$$e^{-iQ\cdot{R}}$ in the $F_1$ layer due to the exchange splitting of the energy bands. This state can be rewritten as a mixture of the spin singlet state and the spin triplet state with zero spin projection on the direction of magnetization: $(\mid\uparrow\downarrow\rangle$$-$$\mid\downarrow\uparrow\rangle)_x$$\cos(QR)$+$i(\mid\uparrow\downarrow\rangle$+$\mid\downarrow\uparrow\rangle)_x$$\sin(QR$). This opposite-spin triplet state $(\mid\uparrow\downarrow\rangle$+$\mid\downarrow\uparrow\rangle)_x$ is equivalent to the equal-spin triplet state $-(\mid\uparrow\uparrow\rangle$$-$$\mid\downarrow\downarrow\rangle)_z$ when viewed with respect to the $z$-axis~\cite{Esc,MatthiasEschrig}. During this process, the magnetization of the $F_1$ layer is preferable to make the spin of one entangled electron align the same direction, then the spin of the other electron will be antiparallel to the $F_1$ direction. The entangled electron inside the $F_1$ ($N$ or $S$) layer can be described as $\mid\uparrow\rangle_{x}{e^{i{k_{\uparrow}}R}}$ ($\mid\downarrow\rangle_{x}{e^{-ik_{\downarrow}R}}$), where $k_{\uparrow(\downarrow)}$=$k_F$+($-$)$Q/2$. The contribution of two entangled electrons to the pair amplitude is proportional to $e^{\pm{i(k_{\uparrow}-k_{\downarrow})R}}$=$e^{\pm{iQR}}$. If the exchange field and the thickness of the $F_1$ layer are weak and short enough, the phase acquired by the spin singlet pair and the opposite-spin triplet pair satisfies the condition $QR<\pi$. The magnetic moment induced in the $N$ and $S$ regions is antiparallel to the magnetization direction of the $F_1$ layer. By contrast, if the thickness $L_1$ or the exchange field $h_1$ becomes large, the above phase is possible to meet the condition $QR>\pi$. In this case, the entangled electrons residing on both sides of the $S/F_1$ interface simultaneously get a $\pi$ phase-shift, and then the $m_x$ direction is reversed. As shown in figures~\ref{Fig.3} (b) and \ref{Fig.3} (d), the pattern configuration of $f_0$ and $f_3$ will be reversed as $h_1/E_F$ increases from 0.035 to 0.1. This changing behavior can demonstrate the results mentioned above. Additionally, as described in reference~\cite{HMeng} the equal-spin triplet pair can be modulated through varying the strength of the exchange field and the thickness of the $F_1$ layer. It can be seen from figure~\ref{Fig.4} that the change of $f_1$ corresponds to the reversal of the $m_z$ direction. Two $m_z$ components decay in the $S$ region and their amplitudes become quite small in the $N$ region. This illustrates the fact that $m_z$ almost cannot be induced in the $N$ region. Taken as a whole, if one changes the strength of the exchange field (or the thickness) of the $F_1$ layer, the direction of total magnetic moment induced in the $S$ region will be reversed. This theoretical result can be used to explain the experimental findings in reference~\cite{DStaNMou}. In $S$/manganite multilayers/$F$ hybrid structure, the manganite multilayers including 15 bilayers [La$_{0.33}$Ca$_{0.67}$MnO$_{3}$/La$_{0.60}$Ca$_{0.40}$MnO$_{3}$]$_{15}$ could offer noncollinear magnetization. Because the thickness of the manganite multilayers is about 120 nm and the total effective exchange field is strong enough, its original properties satisfy the condition $QR>\pi$, in which case the magnetic moment induced in the $S$ region has a fixed direction. Since the applied external magnetic field exceeds the coercive field, the noncollinear magnetization will be reduced and it matches the condition $QR<\pi$, then the induced magnetic moment will change its direction.

   \begin{figure}
   \centerline{\includegraphics[width=6.0in]{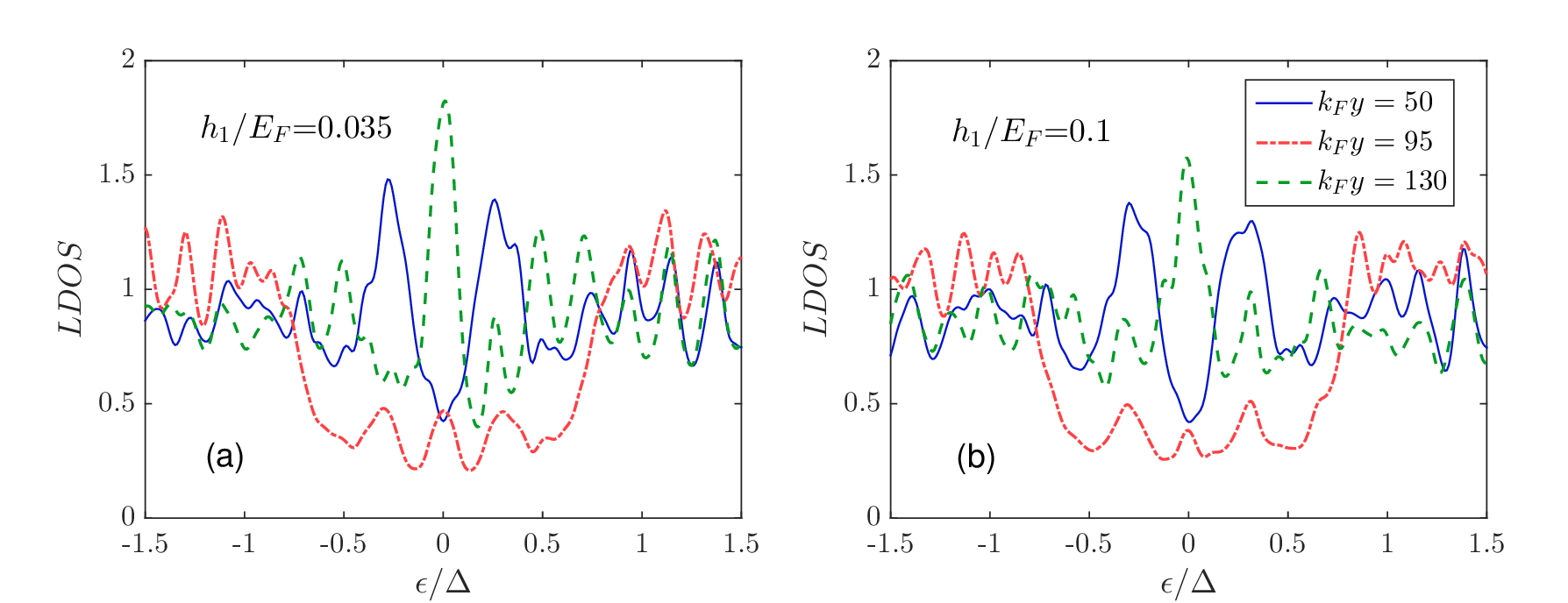}} 
   \caption{The LDOS plotted versus the dimensionless energy $\epsilon/\Delta$ in the $N$ layer ($k_Fy=50$), the $S$ layer ($k_Fy=95$) and the $F_1$ layer ($k_Fy=130$) for two special exchange fields $h_1/E_F=0.035$ (a) and $h_1/E_F=0.1$ (b). The results are calculated at $k_BT=0.001$, and other parameters are the same as in figure~\ref{Fig.3}.}
   \label{Fig.5}
   \end{figure}

    In the experiment, the equal-spin triplet state $f_1$ appearing in the system may be manifested through a zero energy conductance peak (ZECP) in the differential conductance spectrum, in principle, which is proportional to the LDOS. In order to further demonstrate the distribution of the equal-spin triplet pair, we show in figure~\ref{Fig.5} the LDOS in different positions for two special exchange fields $h_1$. In the $F_1$ layer ($k_Fy$=130) there is a sharp ZECP in the LDOS. This ZECP decreases as entering the $S$ layer ($k_Fy$=95). Meanwhile, two subgap peaks appear on both sides of the ZECP, which are caused by the inverse proximity effect of the $N$ layer. In the $N$ layer ($k_Fy$=50), the subgap peaks are enhanced but the ZECP does not appear. This means that the equal-spin triplet pair hardly penetrates into the $N$ layer. Additionally, when the exchange field $h_1$ takes two different values, the variation characteristics of the LDOS are almost identical except that the peaks at $h_1/E_F$=0.035 are higher than that at $h_1/E_F$=0.1.

   \begin{figure}
   \centerline{\includegraphics[width=5.8in]{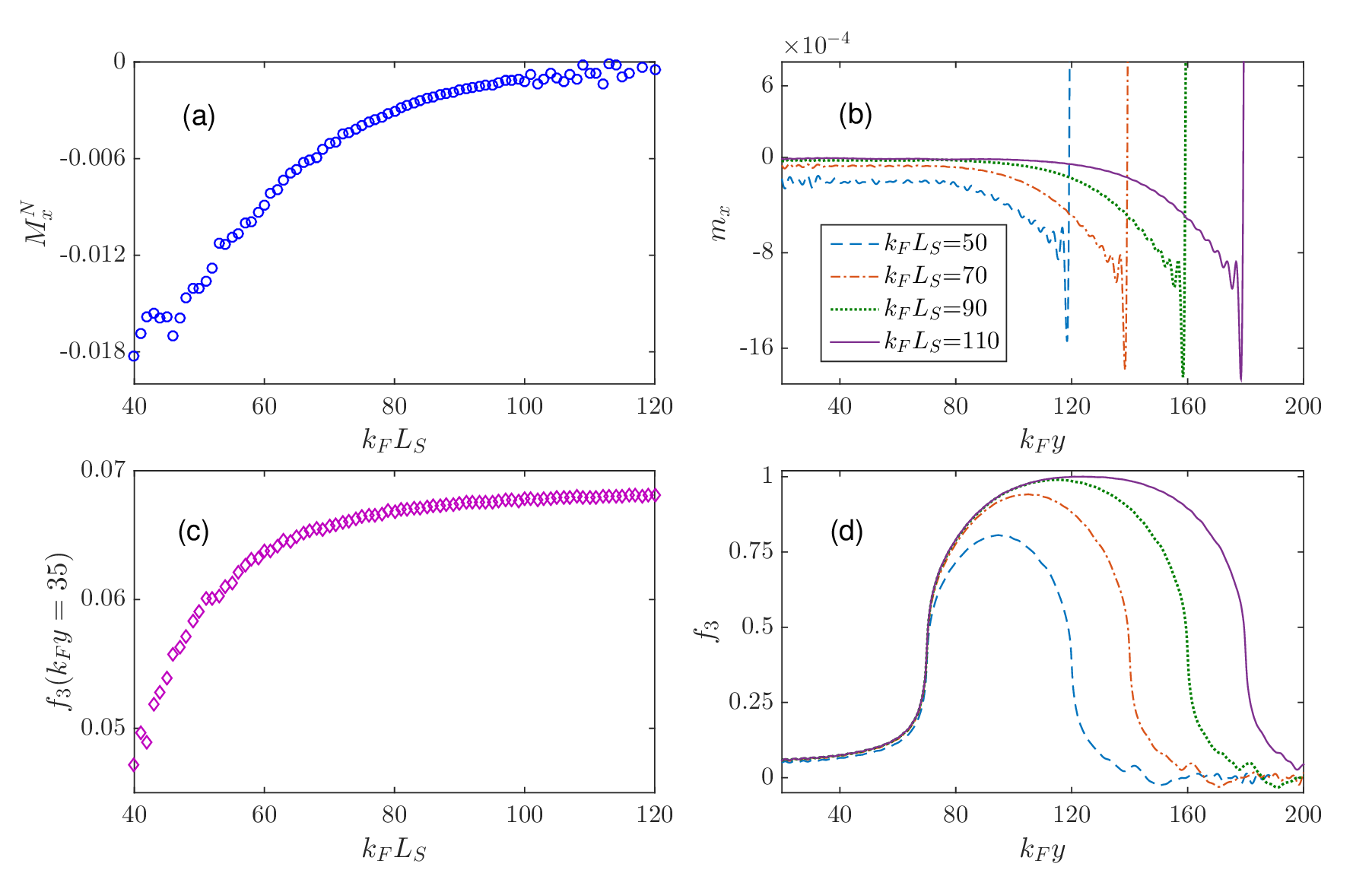}} 
   \caption{The effective magnetic moment $M^N_x$ inside the $N$ region (a) and the spin singlet state $f_3$ in the center of the $N$ layer ($k_Fy=35$) (c) as a function of the superconducting thickness $k_FL_S$. The spatial distribution of the induced magnetic moment $m_x$ (b) and the spin singlet state $f_3$ (d) for the different superconducting thickness $k_FL_S$. The results plotted are for $k_FL_1=20$, $k_FL_2=50$, $h_1/E_F=0.035$, $h_2/E_F=0.2$, $\theta=\pi/2$, and $k_BT$=0. Panels (b) and (d) utilize the same legend.}
   \label{Fig.6}
   \end{figure}

\begin{figure}
   \centerline{\includegraphics[width=6in]{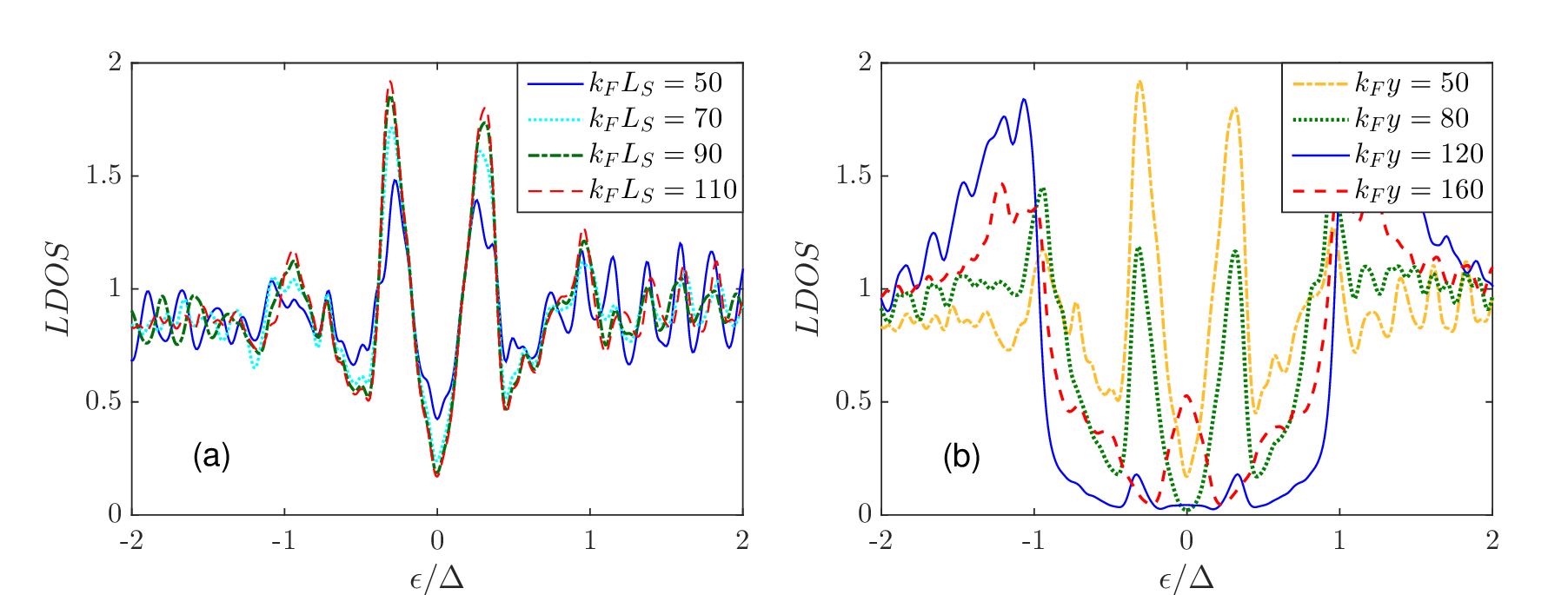}} 
   \caption{(a) The LDOS in the $N$ layer ($k_Fy=50$) plotted versus the dimensionless energy $\epsilon/\Delta$ for different superconducting thicknesses $k_FL_S$. (b) The LDOS at different locations for the fixed superconducting thickness $k_FL_S=110$. The results plotted are for $k_FL_1=20$, $k_FL_2=50$, $h_1/E_F=0.035$, $h_2/E_F=0.2$, $\theta=\pi/2$, and $k_BT=0.001$.}
   \label{Fig.7}
   \end{figure}

    Next, let us consider the effect of the superconducting thickness on the induced magnetic moment. From figures~\ref{Fig.6} (a) and \ref{Fig.6} (b), we can see that the induced magnetic moment inside the $N$ region decreases with an increase of the thickness $L_S$. It is known that the CPS effect only occurs when the superconducting thickness is less than the superconducting coherence length $\xi_{S}$. So the induced magnetic moment disappears under the condition that $k_FL_S$ is larger than $k_F\xi_{S}$=100. By contrast, with the increase of $L_S$ the spin singlet state $f_3$ in the $N$ region will increase (see figures~\ref{Fig.6} (c) and \ref{Fig.6}(d)), which indicates that the increase of the superconducting proximity effect enhances the subgap in the $N$ region. This feature can be demonstrated by the enhancement of the subgap peaks in the LDOS, as illustrated in figure~\ref{Fig.7} (a). The above result shows that the magnetic moment cannot be induced in the $N$ region since two entangled electrons forming the Cooper pair tunnel simultaneously into the $N$ region. Moreover, we show in figure~\ref{Fig.7} (b) the LDOS at different positions when the superconducting thickness is more than the superconducting coherence length. In the $S$ region adjacent to the $F_1$ layer ($k_Fy$=160), there is a distinct ZECP due to the tunneling of the equal-spin triplet pair. Near the center of the $S$ region ($k_Fy$=120) the ZECP vanishes but the subgap peaks appear. As the position shifts from the $S$ region into the $N$ region, in which case $k_Fy$ changes from 120 to 50, the subgap peaks continue to increase until reaching the maximum. In the actual measuring device, similar to the experiment in \cite{MGFNSJKGB}, we propose that Nb and Cu can be selected as possible candidates for superconducting and normal metallic materials, respectively.

   We now turn to discuss the dependence of the effective magnetic moments $M^N_x$ and $M^N_z$ inside the $N$ region on the characteristics of the $F_2$ layer. As shown in figure~\ref{Fig.8}, for the parallel configuration ($\theta$=0) $M^N_x$ does not appear and $M^N_z$ oscillates around the zero value, whose amplitude decays with respect to the external field $h_2$ (or thickness $L_2$). This shows that under the weak exchange field $h_2$ one entangled electron tunnels into the $F_2$ region, and the other electron penetrates into the $N$ region. Since the phase of the entangled electron in the $F_2$ region shifts with $h_2$ (or $L_2$), the phase of the other entangled electron inside the $N$ region changes simultaneously. The phase-shift acquired by the entangled electrons can cause the $M^N_z$ direction to reverse. If the value of $h_2$ (or $L_2$) becomes large, which leads to an increase of the total magnetization of the $F_2$ layer, the tunneling of the entangled electron into the $F_2$ layer will be suppressed, and then $M^N_z$ decreases accordingly. For the perpendicular configuration($\theta$=0.5$\pi$), $M^N_z$ presents the same oscillatory behavior, but $M^N_x$ shows a negative value with superimposed oscillations. These oscillations tend to decrease with increasing $h_2$. The reason may be explained by the following picture. For weak $h_2$ the CPS effect can be divided into two parts: one part occurs between the $N$ and $F_1$ regions and the other part occurs between the $N$ and $F_2$ regions. It should be noted that the CPS effect between the $N$ and $F_2$ regions can induce two components $M^N_z$ and $M^N_x$ inside the $N$ region. The phase of the entangled electron within the $F_2$ layer will shift with increasing $h_2$ (or $L_2$), which results in the oscillation of $M^N_z$ and $M^N_x$. As the strength of $h_2$ (or $L_2$) grows to a higher value, the entangled electron tunneling is suppressed, and then the contribution of the entangled electron to $M^N_x$ and $M^N_z$ decreases. Therefore, $M^N_z$ decreases and $M^N_x$ tends to a constant value. In the actual measurement and application device we recommend that the $F_2$ layer can be chosen to be a strongly spin-polarized ferromagnet, e.g. Co, whose exchange field is 309 meV~\cite{ARobinsonP}, and the $F_1$ layer is taken to be weak one, e.g. Pd$_{90}$Ni$_{10}$, whose exchange field is only 35 meV~\cite{ARobinsonP} and the polarized direction can be modulated easily by the applied magnetic field. In such a case, only the CPS effect between the $N$ and $F_1$ regions still exists, which is beneficial for experimental observation.

   \begin{figure}
   \centerline{\includegraphics[width=6in]{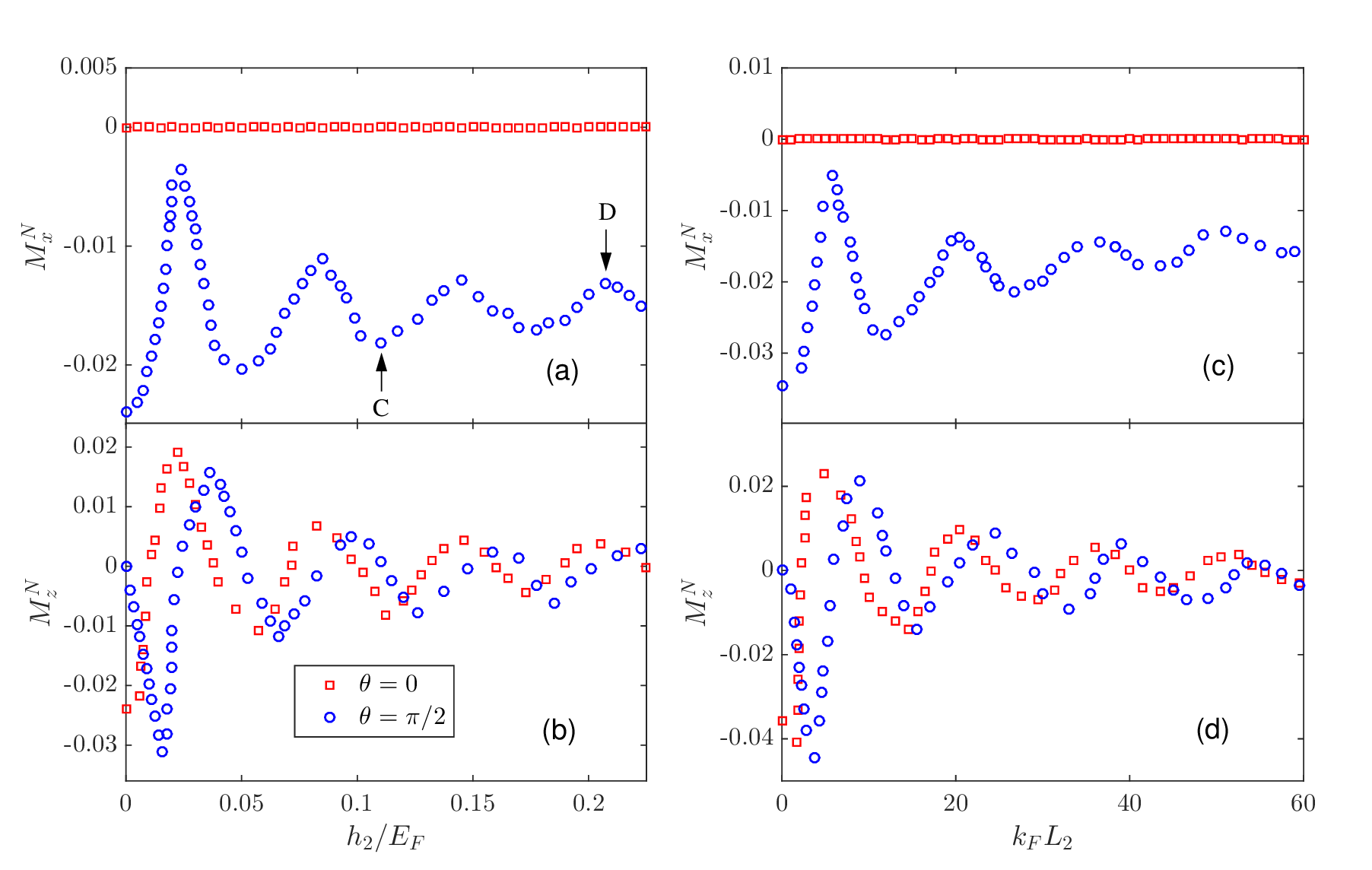}} 
   \caption{The effective magnetic moments $M^N_x$ (a) and $M^N_z$ (b) inside the $N$ region as a function of exchange field $h_2/E_F$ for thickness $k_FL_2=50$. The corresponding $M^N_x$ (c) and $M^N_z$ (d) as a function of thickness $k_FL_2$ for exchange field $h_2/E_F=0.2$. All panels are plotted for $k_FL_S=50$, $k_FL_1=20$, $h_1/E_F=0.035$, and $k_BT=0$.}
   \label{Fig.8}
   \end{figure}

\begin{figure}
   \centerline{\includegraphics[width=6.0in]{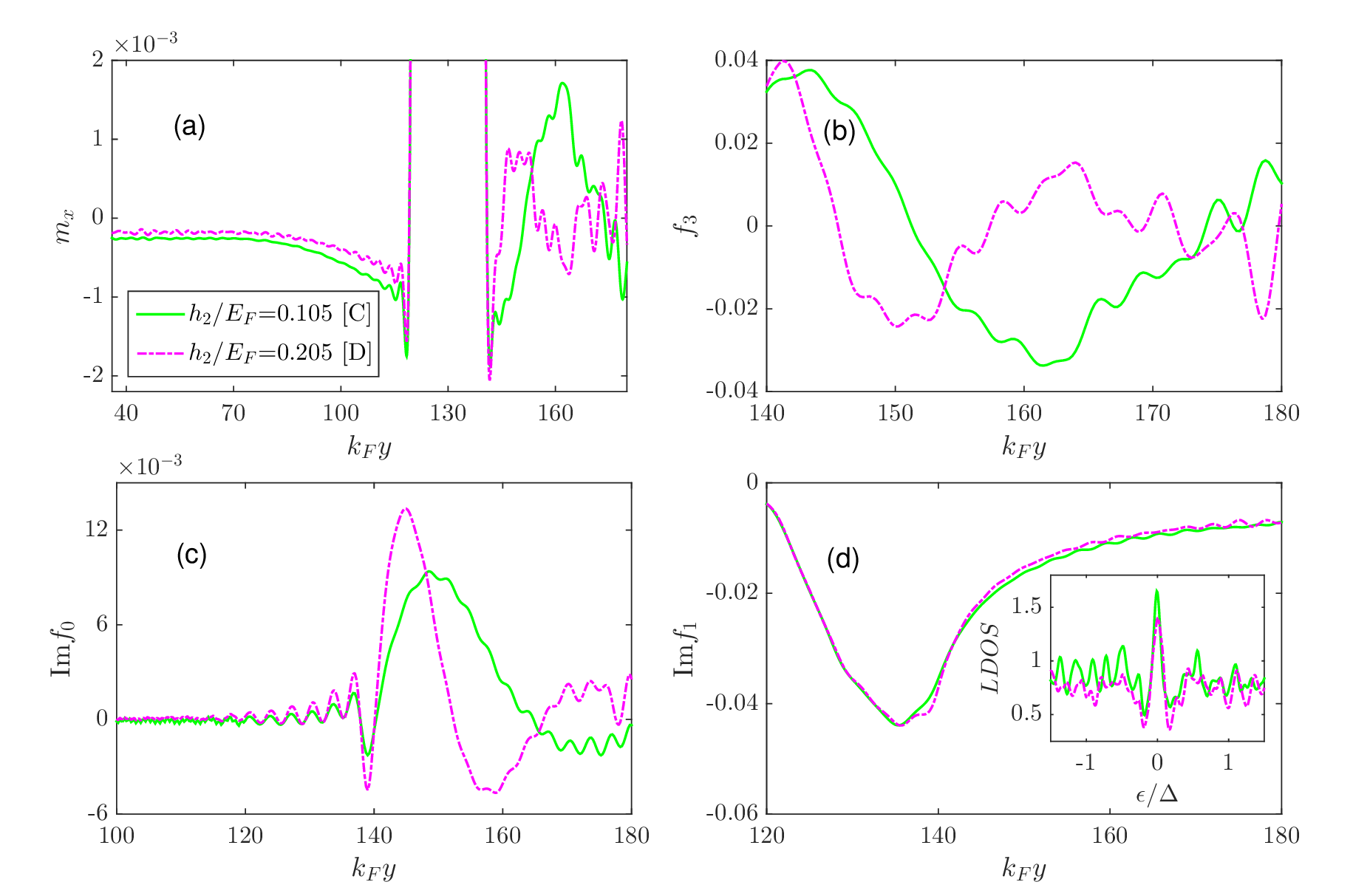}} 
   \caption{The spatial distribution of the induced magnetic moment $m_x$ (a), the spin singlet state $f_3$ (b), and the imaginary parts of the spin triplet states $f_0$ (c) and $f_1$ (d) for two different exchange fields $h_2$. The inset in (d) shows the LODS in the $F_2$ region ($k_Fy=160$), which is calculated at $k_BT=0.001$. Besides, $\omega_Dt=2$ is used in (c) and (d). Other parameters used for the main panels are $k_FL_S=50$, $k_FL_1=20$, $h_1/E_F=0.035$, $k_FL_2=50$, and $k_BT=0$.}
   \label{Fig.9}
   \end{figure}

   The entangled electrons also induce two magnetic components $m_z$ and $m_x$ in the $F_2$ region when they tunnel into this region. The $m_z$ component is superimposed on the original magnetic moment produced by the exchange field $h_2$, and the $m_x$ component is influenced by the exchange field $h_2$. We show in figure~\ref{Fig.9} the spatial profiles of $m_x$, $f_3$ and the imaginary part of $f_0$ and $f_1$ in two cases $h_2/E_F$=0.105 and 0.205, which correspond to two particular points C and D in figure~\ref{Fig.8} (a). We can see that the pattern configurations of $f_3$ and $f_0$ in the $F_2$ region will be reversed when $h_2/E_F$ increases from 0.105 to 0.205. This behavior gives rise to the change of $m_x$ in the $F_2$ region. Meanwhile, we find that $f_1$ changes slightly in the ferromagnetic region, which can be manifested through the ZECP in the LDOS. As mentioned previously $f_1$ may induce the magnetic moment in the $S$ region, but it can not generate an effect in the $N$ region.

\begin{figure}
    \centerline{\includegraphics[width=6in]{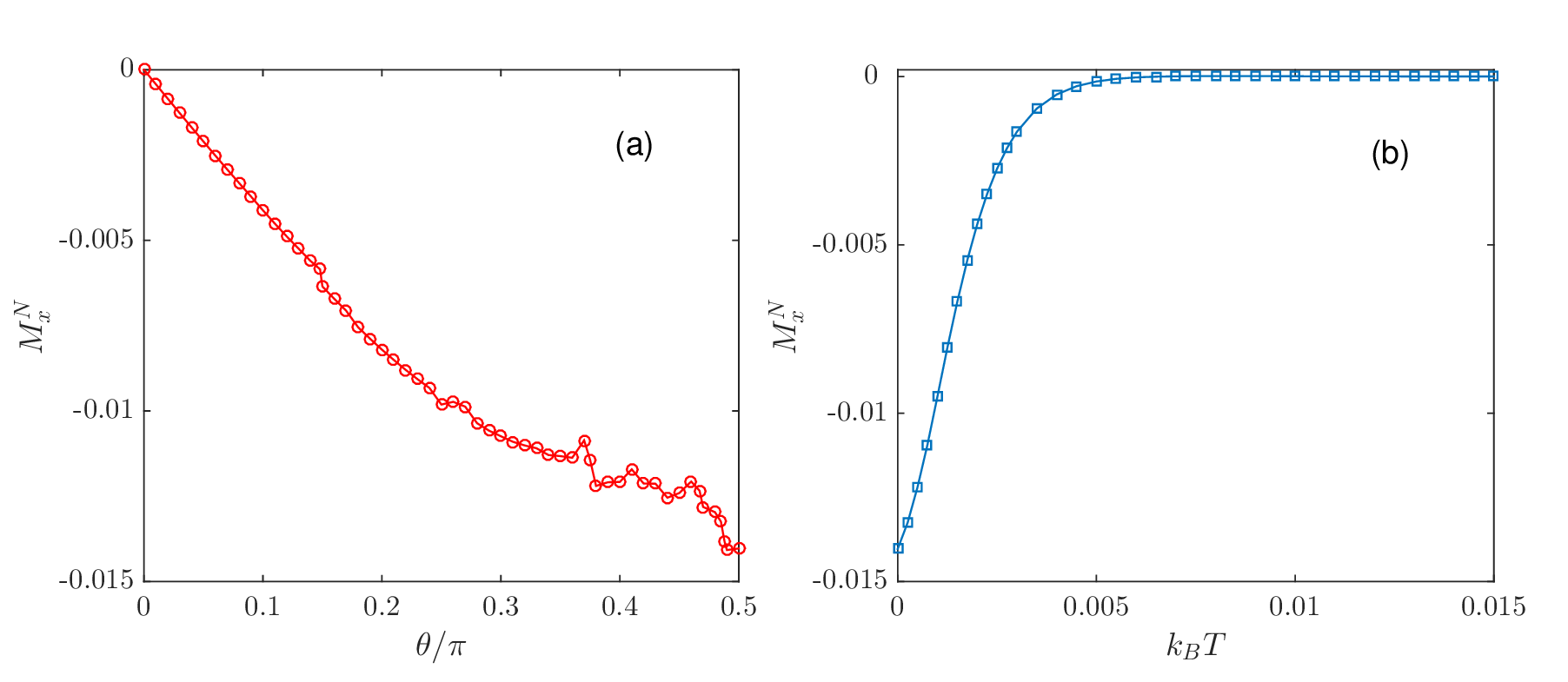}} 
    \caption{(a) The effective magnetic moment $M^N_x$ inside the $N$ region plotted as a function of the misorientation angle $\theta$ at temperature $k_BT=0$. (b) $M^N_x$ as a function of temperature $k_BT$ in the orthogonal arrangement ($\theta=0.5\pi$). Parameters used in all panels are $k_FL_S=50$, $k_FL_1=20$, $h_1/E_F=0.035$, $k_FL_2=50$, and $h_2/E_F=0.2$.}
   \label{Fig.10}
   \end{figure}

   The dependence of the effective magnetic moment $M^N_x$ inside the $N$ region on the misorientation angle $\theta$ and the temperature $T$ is represented in figures~\ref{Fig.10}. It is possible to find that $M^N_x$ turns from zero to a negative value as the two exchange fields switch from parallel to orthogonal polarization, which means that $M^N_x$ is directly related to $x$-component of magnetization in the $F_1$ layer. This effect may be used for engineering cryoelectronic devices to manipulate the induced magnetic moment in the $N$ region. Moreover, $M^N_x$ decreases rapidly with increasing temperature. Particularly, if the temperature rises high enough, $M^N_x$ completely disappears. This demonstrates that $M^N_x$ arises from the entangled electron tunnelling into the $N$ region but is not caused by the magnetization leakage from the $F_1$ region into the $N$ region.

   We compare our theoretical result with Flokstra \emph{et al}'s experiment~\cite{MGFNSJKGB}. Our result shows that the magnetic moment could be induced in both the $S$ and $N$ regions. However, in the experiment~\cite{MGFNSJKGB} the magnetic moment does not appear in the $S$ region but only occurs in the adjacent $N$ region. Here we propose two kinds of conjectures to resolve this contradiction. The first possibility is that the magnetic moment induced in the $S$ region is suppressed by other mechanisms, such as the spin-orbit interaction and orbital effects (Meissner currents)~\cite{AAALPG,FSBAFVolkov}. From figure 2b in reference~\cite{MGFNSJKGB}, one can see that a portion of the magnetic moment still presents in the $S$ region although its amplitude is relatively small. We speculate that the accumulation of the entangled electrons will induce a faint Zeeman field in the superconducting region. The flux caused by the Zeeman field will be expelled by the Meissner effect. So the induced magnetic moment can not be measured in the experiment~\cite{MGFNSJKGB}. Very recently, an anomalous Meissner screening has been observed in normal-metal/superconductor (Cu/Nb) and normal-metal/superconductor/ferromagnet (Cu/Nb/Co) thin films~\cite{FMGSR}. Comparing with isolated superconductor (50-nm-thick Nb film), the flux expulsion becomes significantly enhanced when adding an adjacent normal-metal (40nm Cu layer). This indicates that the added normal-metal is effective to help the superconductor expel flux. Moreover, a further significant enhancement of the flux expulsion is observed when adding a ferromagnet (2.4nm Co layer) to the other side of the superconductor (Nb). From this experiment, we can infer that the magnetic flux expulsion in the $NSF$ is stronger than that in the $SF$. So the experimental results in reference~\cite{FMGSR} may account for the fact that the magnetic moment induced in the $S$ region can be observed in the $SF_1F_2$~\cite{GAOvsyannikov} but not in the $NSF_1F_2$~\cite{MGFNSJKGB}. The second possibility is that the direction of the magnetic moment induced in the $S$ region deviates from that in the $N$ region, hence the magnetic moment inside the $S$ region is not easily measured in the experiment. As mentioned before, the magnetic moment along the $x$-axis is simultaneously induced in both the $N$ and $S$ regions by the CPS effect, but the magnetic moment along the $z$-axis is only produced in the $S$ region by the equal-spin triplet pairs. Therefore, the direction of the total magnetic moment in the $S$ region deviates from the $x$-axis. On the other hand, there are two same effects in our result and Flokstra \emph{et al}'s experiment~\cite{MGFNSJKGB}: (i) The induced magnetic moment in the $N$ region exhibits a spin-valve effect, which means that the magnetic moment depends on the mutual orientation of the two ferromagnetic layers. (ii) The induced magnetic moment can be controlled by temperature. Below superconducting critical temperature $T_c$, the magnetic moment appears in the $N$ region for the orthogonal arrangement, but above $T_c$ the corresponding magnetic moment disappears.

   Finally, we give a brief discussion on the disorder effect. In the present work, the clean limit is taken by considering that the characteristic size of the total $NSF_1F_2$ structure is smaller than the electron mean free path. The good agreement between our recent work~\cite{HMeng} and experiments~\cite{TruptiSKhaire,CarolinKlose} indicate that the clean limit used here is a good approximation, and the disorder due to impurity scattering and interface roughness is of no importance to the changes in the physical characteristics. In our calculation, the non-local electron tunneling plays an important role in the CPS effect. With the increase in barrier strength at the $N$/$S$, $S$/$F_1$, and $F_1/F_2$ interfaces, the transportation of the entangled electrons will be suppressed. For the same reason, the disorder also gives rise to a reduction in the CPS effect, which is unfavorable to the appearance of the induced magnetic moment.

   \section{Conclusion}

   In this article, we have investigated the magnetic moment induced in the $N$ and $S$ regions in the $NSF_1F_2$ junctions with the misorientation magnetization. We find that the induced magnetic moment arises from the spin-entangled electrons accumulating in these regions. In contrast, the equal-spin triplet pair within the $S$ region also can induce a magnetic moment in the different direction, but this magnetic moment hardly penetrates from the $S$ region into the $N$ region. Moreover, the variation of the exchange field and the thickness of the $F_1$ layer leads to the reversal of the induced magnetic moment inside the $N$ region. This behavior is attributed to the phase-shift acquired by the spin-entangled electrons tunneling into the $F_1$ layer. On the other hand, the change of the $F_2$ layer causes the oscillation of the induced magnetic moment, and this oscillating character will be suppressed as the increase of the exchange field in the $F_2$ layer. In addition, the induced magnetic moment in the $N$ region increases with misorientation angle and reaches a maximum for the orthogonal arrangement, but it decreases with increasing temperature. Thus, all these findings may provide new insight into the physical mechanism for the induced magnetic moment inside the $N$ and $S$ regions.

   \ack
   This work was supported by the National Natural Science Foundation of China (Grants No.11604195 and No.11447112), National natural science foundation for theoretical physics special fund (Grant No.11547039), Youth Hundred Talents Programme of Shaanxi Province, and the Opening Project of Shanghai Key Laboratory of High Temperature Superconductors (Grant No.14DZ2260700). J. Wu was supported by National Natural Science Foundation of China (Grants No.11674152 and No.11681240276), Guangdong Innovative and Entrepreneurial Research Team Program (No.2016ZT06D348) and Natural Science Foundation of Guangdong Province (Grant No.2017B030308003) and Science, Technology and Innovation Commission of Shenzhen Municipality (Grants No.ZDSYS201703031659262 and No.JCYJ20150630145302225) and Shenzhen Basic Research Fund (Grant No.JCYJ20170412152620376) and the startup funding of the South University of Science and Technology of China and the Shenzhen Peacock Plan.

    \end{document}